\documentclass[aps,prb,reprint,groupedaddress,amsmath,showpacs]{revtex4-1}
\usepackage{graphicx}
\usepackage{amssymb}
\usepackage{units}
\usepackage{booktabs}
\usepackage{dcolumn}

\begin{document}
\title{Low-temperature ordered phases of the spin-$\frac{1}{2}$ XXZ chain system Cs$_2$CoCl$_4$}
\author{O. Breunig$^1$}

\author{M. Garst$^2$}
\author{A. Rosch$^2$}
\author{E. Sela$^{2,3}$}
\author{B. Buldmann$^2$}

\author{P. Becker$^4$}
\author{L. Bohat\'y$^4$}
\author{R. M\"uller$^1$}

\author{T. Lorenz$^1$}

\affiliation{$^{1}$II. Physikalisches Institut, Universit\"at zu K\"oln, Z\"ulpicher Str. 77, 50937 K\"oln, Germany}
\affiliation{$^{2}$Institut f\"ur Theoretische Physik, Universit\"at zu K\"oln, Z\"ulpicher Str. 77, 50937 K\"oln, Germany}
\affiliation{$^{3}$Raymond and Beverly Sackler School of Physics and Astronomy, Tel-Aviv University, Tel Aviv 69978, Israel}
\affiliation{$^{4}$Institut f\"ur Kristallographie, Universit\"at zu K\"oln, Greinstra\ss{}e 6, 50939 K\"oln, Germany}
\date{\today}

\begin{abstract}
In this study the magnetic order of the spin-1/2 XXZ chain system Cs$_2$CoCl$_4$ in a temperature range from \unit[50]{mK} to \unit[0.5]{K} and in applied magnetic fields up to \unit[3.5]{T} is investigated by high-resolution measurements of the thermal expansion and the specific heat. Applying magnetic fields along \textit{a} or \textit{c} suppresses $T_\textrm{N}$ completely at about \unit[2.1]{T}. In addition, we find an adjacent intermediate phase before the magnetization saturates close to 2.5 T. For magnetic fields applied along \textit{b}, a surprisingly rich phase diagram arises. Two additional transitions are observed at critical fields $\mu_0 H_{SF1}\simeq\unit[0.25]{T}$ and $\mu_0 H_{SF2}\simeq\unit[0.7]{T}$, which we propose to arise from a two-stage spin-flop transition.
\end{abstract}

\maketitle

\section{Introduction}
The magnetism of low-dimensional spin systems is of fundamental interest due to its relation to many recent problems of modern solid-state physics, \textit{e.g.} high-temperature superconductivity, the emergence of exotic ground states or quantum critical phenomena at low temperatures~\cite{Sachdev2011}. In this context, one-dimensional systems are especially appealing due to the enhanced quantum fluctuations. Some of the applied theoretical models are even exactly solvable, \textit{e.g.} the transverse-field Ising model that was successfully applied to the famous compounds LiHoF$_4$ \cite{Bitko1996} and CoNb$_2$O$_6$ \cite{Coldea2010}, which may serve as model magnets for field-induced quantum criticality. Most studies of one-dimensional magnets focus on the temperature range, in which the physics is governed by the primary magnetic exchange. The emergence of long-range magnetic order at low temperatures is not covered by the applied one-dimensional models, which, \textit{e.g.}, based on the Mermin-Wagner theorem do not show magnetic order at any finite temperature. Nevertheless, magnetic order does arise in corresponding model crystals due to finite interchain couplings. Although these couplings might be comparably small, their impact on the low-energy excitations in the vicinity of quantum critical points can lead to interesting physics like the emergence of a unique symmetry in the compound CoNb$_2$O$_6$ \cite{Coldea2010}.

\begin{figure}[t!]
  \centering
  \includegraphics[width=\columnwidth]{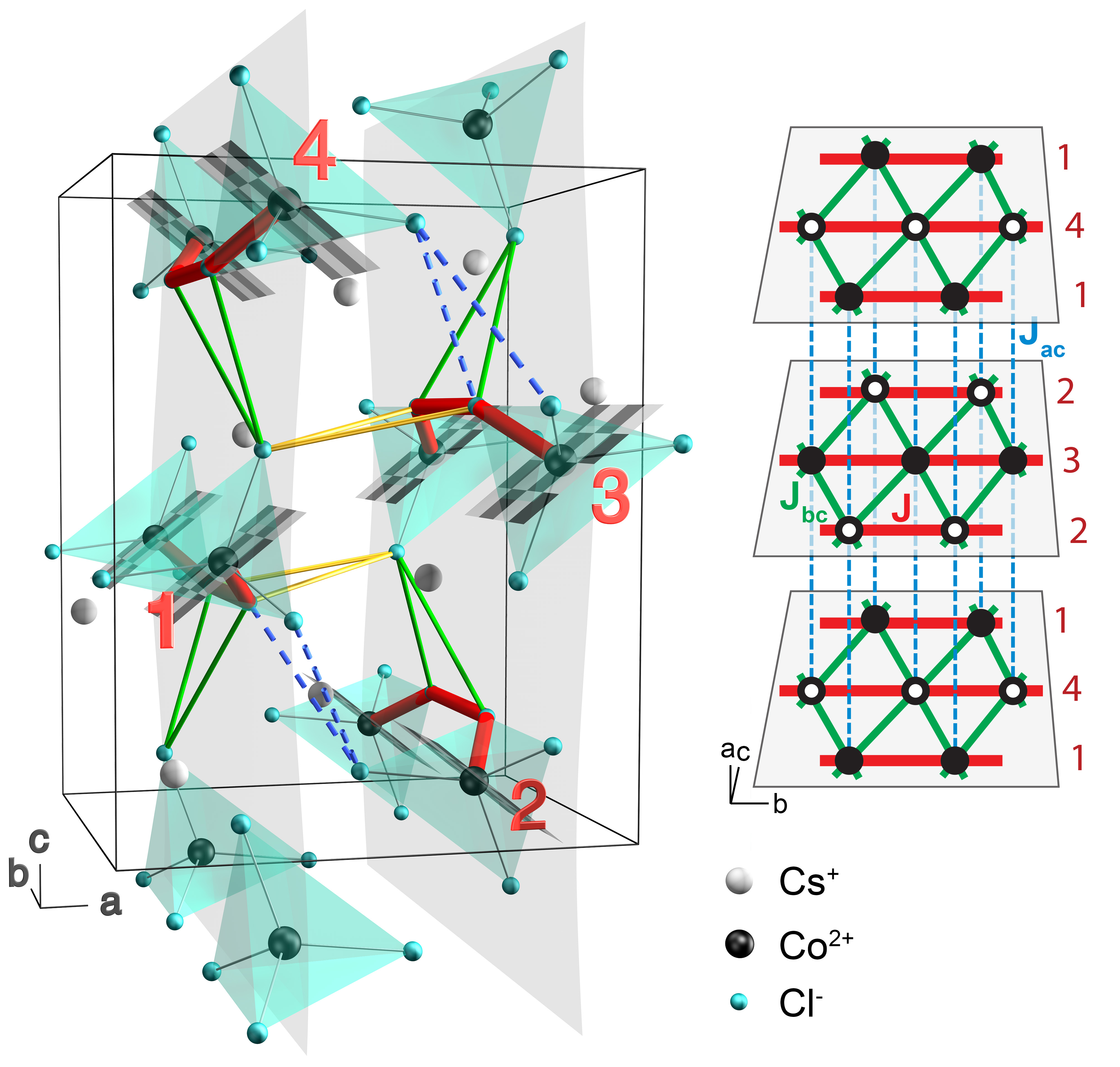}
 \caption{(color online) Crystal structure of Cs$_2$CoCl$_4$ (left) based on  Ref.~\onlinecite{Figgis1987}, but with the unit cell shifted by (0,0.5,0.5) as in Ref.~\onlinecite{Kenzelmann2002}. Checkerboard planes indicate the orientations of the  magnetic easy planes alternating between sites (1\&3) and (2\&4). The dominant intrachain exchange $J$ is along \textit{b} via Co-Cl-Cl-Co paths (sketched in red). Along \textit{c}, the chains form buckled layers (grey) with frustrated interchain couplings $J_{bc}$ (green) between 1(2) and 4(3). Along the stacking direction \textit{a}, there are non-frustrated interlayer couplings $J_{ac}$ (blue) between 1(3) and 2(4) and also frustrated couplings  $J_{ab}$ (yellow) between 1(2) and 3(4), but the latter are  only present between every second pair of \textit{bc} layers. On the right is a schematic representation of the magnetic lattice with equivalently colored couplings $J$, $J_{bc}$, $J_{ac}$, but without $J_{ab}$. The alternating easy-plane orientations are represented by open and filled symbols.}
  \label{fig:struktur}
 \end{figure}
 
An interesting class of models, where quantum critical phenomena can be observed in thermodynamic properties, are models with a planar anisotropy of the XXZ type, when the rotational symmetry around the $z$ axis is broken by a magnetic field applied in the transverse direction, \textit{i.e.} perpendicular to the anisotropy axis. At temperatures below \unit[2]{K}, Cs$_2$CoCl$_4$ is an established realization of this type \cite{Algra1976,smit1979field,mukherjee2004field,Breunig2013}. In this compound the magnetism arises from Co$^{2+}$ ions that are surrounded by distorted Cl$_4$ tetrahedra and form chains along the crystallographic \textit{b} axis. Due to the induced strong crystal-field anisotropy $D$ the orbital $S=3/2$ quartet is split into two Kramers doublets that are separated by $\Delta E=2D\simeq 14\,\text{K}$. Thus, at low temperatures $T\ll\Delta E$, a description of the lower $\left|\pm\frac{1}{2}\rangle\right.$ states in terms of an effective spin-1/2 model arises. The primary magnetic exchange between
CoCl$_4$ tetrahedra is found along \textit{b}, while inter-chain interactions are estimated to be at least one order of magnitude smaller\cite{McElearney1977,Yoshizawa1983a}.

The anisotropy $D$ establishes an easy-plane anisotropy of the Co$^{2+}$ magnetism. Due to the presence of $2_1$-screw axes in the space group \textit{Pnma} ($D^{16}_{2h}$, No. 62) two equivalent types of tetrahedral coordinations of the cobalt sites arise (cf.~fig.~\ref{fig:struktur}). They give rise to two types of magnetic easy planes that only differ by the sign of a rotation around \textit{b}, that alternates along \textit{c}, but not along the chain direction \textit{b}. This renders the crystallographic \textit{b} axis the only principal axis with respect to the easy-plane orientation. Thus, applying a magnetic field along this axis leads to a description of a single chain in Cs$_2$CoCl$_4$ in terms of the XXZ model in transverse magnetic field
\begin{align}
\label{eqn:HXXZ}
\nonumber\mathcal{H}_{\rm XXZ} =\sum_i J \Big[ &  (S_i^x S_{i+1}^x+S_i^y S_{i+1}^y ) + \Delta S_i^z S_{i+1}^z \\
 &- g \mu_0 \mu_\mathrm{B} H_b S_i^y \Big].
\end{align}

Here, the spin component $y$ is identified by the crystallographic \textit{b} axis, while $z$ defines the local easy plane which alternates from one chain to another. In a previous work \cite{Breunig2013}, we showed that this model is applicable to Cs$_2$CoCl$_4$ in a temperature range from 0.25 K to 2 K and in magnetic fields up to 3 T. The anisotropy $\Delta\simeq 0.12$ was determined from an analysis of specific-heat and thermal-expansion data, which furthermore show clear signatures of quantum criticality close to \unit[2]{T}.

Cs$_2$CoCl$_4$ is isostructural to the intensely studied compound Cs$_2$CuCl$_4$, which shows diverse magnetic phases at low temperatures \cite{Coldea2001,Radu2005,Starykh2010}. Identifying the couplings of Cs$_2$CoCl$_4$ in analogy to Ref.~\onlinecite{Kenzelmann2002} leads to a representation of the magnetic lattice, which is identical to that of Cs$_2$CuCl$_4$ (right of fig.~\ref{fig:struktur}). It consists of anisotropic triangular layers within \textit{bc} planes that are stacked along \textit{a}. In Cs$_2$CuCl$_4$ the exchange constants $J_{bc}$ and $J$ within the triangular layers are of comparable magnitudes~\cite{Coldea2002}. In contrast,  the interchain interactions of Cs$_2$CoCl$_4$ were estimated to be at least one order of magnitude smaller than the dominant intrachain interaction $J$ \cite{Yoshizawa1983a,Chatterjee2003}. Thus,  Cs$_2$CoCl$_4$ is close to the spin-chain limit and magnetic order is observed at lower temperatures than in Cs$_2$CuCl$_4$.
Another important difference between both compounds arises from the different electronic configurations of copper and cobalt. In case of Cu$^{2+}$($3d^9$) the orbital momentum is quenched by the crystal electric field, which leads to an almost fully isotropic Heisenberg magnetism. In contrast, the orbital momentum of Co$^{2+}$($3d^7$) is finite and spin-orbit coupling may cause strongly anisotropic magnetic properties. In general, the type of anisotropy, XY versus Ising, depends on the coordination and in Cs$_2$CoCl$_4$ magnetic easy planes emerge, which are close to the pure XY limit~\citep{Breunig2013}. As the leading interchain couplings are between sites of different easy-plane orientations (see fig.~\ref{fig:struktur}), we suggest that different mechanisms are relevant for the magnetic order than in Cs$_2$CuCl$_4$.

The antiferromagnetic order of Cs$_2$CoCl$_4$ at temperatures $T<T_\textrm{N}\simeq\unit[220]{mK}$ was previously investigated by neutron scattering\cite{Kenzelmann2002}. At zero magnetic field the spins along each chain order antiferromagnetically with collinear spins oriented within the \textit{bc} planes and tilted away from the \textit{b} axis by $|\phi|\approx 15^\circ$. The sign of $\phi$ is equal for neighboring chains that are coupled via the non-frustrated $J_{ac}$, while the sign of $\phi$ alternates between chains coupled via the frustrated coupling $J_{bc}$. Thus, the collinear spins of chains (1 \& 2) are canted with respect to those of chains (3 \& 4) and all spins are tilted away from their magnetic easy planes. Note that \textit{b} is the common axis of both types of easy-plane orientations and an alternating orientation of the spins along $\pm b$ would yield a collinear antiferromagnetic spin structure with all the spins oriented within their respective magnetic easy planes.  The fact that this most simple N\'{e}el state is not realized in  Cs$_2$CoCl$_4$ reveals that additional couplings, \textit{e.g.}\ between further neighbors or antisymmetric Dzyaloshinskii-Moriya (DM) exchange, have to be considered to understand the complex magnetic structure of this material. This aspect as well as the competition between an incommensurate spin-spiral state favored by the frustrated interchain coupling $J_{bc}$ and the presence of alternating easy-plane orientations have been raised already in Ref.~\onlinecite{Kenzelmann2002}, but have not been clarified until now. Concerning the influence of a magnetic field, it has been found that $T_\textrm{N}$ is fully suppressed around $\unit[2.1]{T}$ for $H\| a$, while the magnetization saturates at slightly larger magnetic fields $\mu_0 H_m\simeq\unit[2.5]{T}$, which was supposed to indicate a spin-liquid ground state in the intermediate field range\cite{Kenzelmann2002}. Studies of the magnetic order of Cs$_2$CoCl$_4$ for  magnetic field directions other than $a$ are, however, not available. Of particular interest is the influence of magnetic fields applied along $b$, as this direction is common to both types of easy planes. Moreover, this field direction is almost collinear to the ordered moments and is thus expected to induce spin-flop transitions. 

Extending our previous work on the one-dimensional magnetism \cite{Breunig2013}, in the present study, we strive for a description of the thermodynamics and the magnetic phases of Cs$_2$CoCl$_4$ at low temperatures and in magnetic fields applied along different axes. We discuss the phase diagrams at temperatures $T<T_\text{N}$ obtained by measurements of specific heat and thermal expansion in magnetic fields applied along different crystallographic axes and discuss possible origins of the observed phases.

\section{Experiment}

Single crystals of optical quality of Cs$_2$CoCl$_4$ were grown from an aqueous solution with a stoichiometric ratio of $1 : 2$ of the educts CoCl$_2\cdot$6H$_2$O and CsCl by controlled evaporation of the solvent at \unit{311}K. During a growth period of 6--8 weeks crystals of dimensions up to 20$\times$20$\times$15\,mm$^3$ with well developed morphology were obtained. The morphological faces of the crystals and X-ray diffraction were used for the sample orientation. Using an inside-hole saw oriented samples of typical dimensions of 2$\times$2$\times$1\,mm$^3$ were prepared. All measurements were performed in a high-vacuum chamber of a dilution refrigerator (Kelvinox 300, Oxford Instruments).

The heat capacity was obtained by the quasi-adiabatic heat-pulse method using a home-built calorimeter, which was previously calibrated in applied magnetic fields. The sample was fixed to the sample platform by a small amount of Apiezon N grease. The addenda heat capacity was measured in a separate run and subtracted.

At temperatures below $\unit[0.5]{K}$ the obtained raw data suffer from an increasing internal relaxation time ($\tau_2$). As different sample shapes as well as surface preparation did not significantly influence the internal relaxation, we conclude that it arises from a small thermal conductivity of Cs$_2$CoCl$_4$ at low temperatures. In the process of thermal equilibration after applying a heat pulse to the sample, parts of the sample pick up more heat than others. In the vicinity of phase transitions, this may lead to heating only parts of the sample above the transition temperature. Thus, the subsequent temperature relaxation is non-trivial. In few of our zero-field raw data, this effect causes a step-like temperature relaxation, indicating latent heat in parts of the crystal. These effects complicate the exact determination of the heat capacity. A secondary effect is the suppression of a temperature hysteresis of $c_p$ due to the partial crossing of $T_\text{N}$ while applying a heat pulse. Therefore the absolute values of $c_p$ close to $T_\textrm{N}$ are not reliable. In the low-field range ($<\unit[0.2]{T}$) the transition temperature is fixed by only few data points. Although the peak shape is therefore not fully resolvable, we interpret the appearance of latent heat as an indication for
the first-order character of these transitions. In the analysis of the heat pulses, $\tau_2$ effects were accounted for as described in Ref.~\onlinecite{Shepherd1985}. Nevertheless, we estimate the induced systematic error to be of the order of $10\%$. However, none of our conclusions is tampered by this effect. Especially the obtained transition temperatures and fields are hardly influenced by the present $\tau_2$ effect.

Thermal expansion and magnetostriction were measured on a home-built capacitance dilatometer made of copper. The capacitance was measured with an AC capacitance bridge (AH2550A, Andeen Hagerling). To ensure that the obtained data are not influenced by magnetocaloric effects, sweeps were performed at very slow magnetic-field rates down to \unit[1]{mT/min}. All presented data were obtained in a longitudinal configuration, meaning that the magnetic field was always applied along the same axis $i$ the relative length change $\Delta L$ was measured for. The uniaxial thermal-expansion coefficient $\alpha_i$ and the magnetostriction coefficient $\lambda_i$ were obtained numerically, $(\alpha_i,\lambda_i)=\frac{1}{L_i}¸\frac{\partial\Delta L_i}{\partial (T,\mu_0 H)}$.

\section{Results}
\subsection{Magnetic field along a or c}
\begin{figure}[tb]
  \includegraphics[width=\columnwidth]{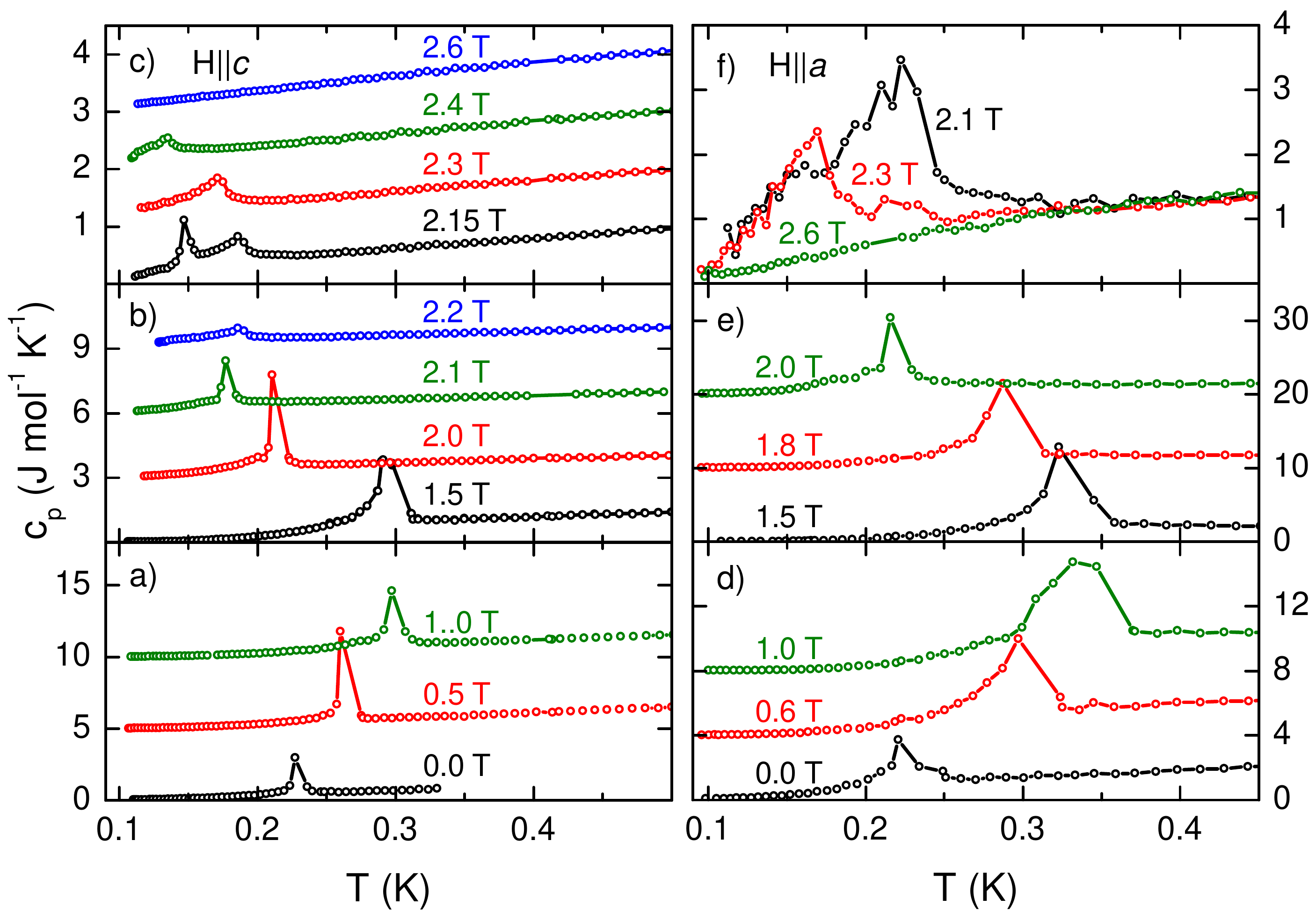}
  \caption{(Color online) Specific heat of Cs$_2$CoCl$_4$ for representative magnetic fields applied along \textit{c} (left) or \textit{a} (right). Curves are offset with respect to each other by 5, 3, 1, 4, and 10~Jmol$^{-1}$K$^{-1}$ in panels a) -- e), respectively.}
  \label{fig:cpHaHc}
 \end{figure}

  The specific heat of Cs$_2$CoCl$_4$ in magnetic fields up to 2.6 T applied along \textit{a} or \textit{c} is shown in fig.~\ref{fig:cpHaHc}. In zero magnetic field, the transition upon cooling from the paramagnetic to the antiferromagnetic phase is reflected by a sharp anomaly at $T_\textrm{N}= 220 \pm 5$ mK. Similar transition temperatures were found in previous studies \cite{Algra1976,Kenzelmann2002}. We observe a linear temperature dependence of the molar specific heat $c_p/N_\text{A} k_\text{B} \simeq 0.43\, T$ for $T_\textrm{N}<T<0.5$ K, as expected in the low-temperature limit ($T\ll J$) of the antiferromagnetic spin-chain model with an easy-plane type anisotropy,
  \begin{equation}
	\frac{C}{R}= A \frac{k_\text{B} T}{J},
  \end{equation}
 with the molar gas constant $R=N_\text{A} k_\text{B}$. The value of the slope $A=\frac{2}{3} \frac{\arccos(\Delta)}{\sqrt{1-\Delta^2}}$ depends on the anisotropy $\Delta<1$ and ranges from
 $A_{xy}=\pi/3 \approx 1.05$ in case of the XY model ($\Delta=0$) to $A_\text{H}=2/3$ for the Heisenberg model\cite{Giamarchi2004}. In the present case of $\Delta\simeq 0.12$ we obtain $J/k_\text{B}\simeq \unit[2.3]{K}$ from our data. Keeping in mind, that the limit of $T\ll J$ is only partially fulfilled, this value compares well to $J/k_\text{B}\simeq\unit[2.9]{K}$ derived from more extensive comparisons at elevated temperatures \cite{Breunig2013, Algra1976}.

  Within the magnetically ordered phase the temperature dependence of the specific heat is well described by a power law $c_p\propto T^\alpha$ with $\alpha\simeq 3.5$. As the phonon heat capacity can be neglected in the present temperature range, the power-law dependence stems from the ordered magnetic subsystem. In the simplest case of an ordered antiferromagnet, one expects a temperature dependence of specific heat $c_p\propto T^{d/n}$, where $d$ is the dimension of the system and $n$ the leading exponent of the magnetic dispersion relation $\omega(\vec k)\propto k^n$. However, the expected $T^3$ dependence of an antiferromagnet ($d=3$, $n=1$) is not found in the experiment. The observed value of $\alpha=3.5$ is probably related to anisotropies $\gamma$ with respect to the chain direction $y$ and the appearance of an anisotropy gap $\mathcal{D_A}$ in the magnon dispersion
  \begin{equation}
  \epsilon(\vec k)=\sqrt{\left[2 J S  \left(\sin k_y +  \gamma (\sin k_x+\sin k_z)\right)\right]^2 + \mathcal{D_A}^2}.
  \end{equation}
  The combination of $\gamma$ and $\mathcal{D_A}$ opens a temperature range in which $c_p$ is well approximated by $T^\alpha$. Comparing numerical calculations of the specific heat with our data, we find a quantitative agreement below $T_\text{N}$ using the parameters
  \begin{equation}
  J/k_\text{B}=\unit[2.9]{K} \quad  \gamma=0.05 \quad \mathcal{D_A}/k_\text{B}=\unit[0.7]{K}.
  \end{equation}
  Here, $J$ was fixed to the established high-temperature value and the relative strength of the inter-chain coupling $\gamma$ was inferred from neutron data \cite{Yoshizawa1983a}. However, we suggest not to stress the exact numerical results too much, as the model allows for a rather broad range of parameters that lead to similar descriptions of the data. Our comparison shows that the observed power law might be well explained by an anisotropic magnon dispersion, which effectively generates a power-law dependence of $c_p$ in a restricted temperature range. However, these results ask for a detailed study by, \textit{e.g.}, electron spin resonance or inelastic neutron scattering to quantify the size of $\mathcal{D_A}$.

  Applying magnetic fields up to 1.0 T, the temperature dependence of $c_p$ remains well described by the same power law $c_p\propto T^{3.5}$. Besides, the magnetic-order peak is shifted to higher temperatures with increasing magnetic field. In case of $H\parallel\textit{a}$, a maximum peak position $T_\textrm{N}\simeq\unit[330]{mK}$ is reached in a magnetic field of \unit[1.0]{T}. Magnetic fields applied along \textit{c} as well lead to an enhancement of the transition temperature to $\simeq\unit[300]{mK}$ at \unit[1.0]{T}. Increasing the magnetic field further (center panels of fig.~\ref{fig:cpHaHc}), the peak position shifts back to lower temperatures. Almost no field dependence is observed, however, in a small adjacent field range ($\unit[2.0]{T}<\mu_0 H\parallel\textit{a}<\unit[2.2]{T}$). In fact, in case of $\unit[2.1]{T}<\mu_0 H\parallel\textit{c}<\unit[2.3]{T}$ even a slight enhancement of $T_\textrm{N}$ can be resolved due to the higher data quality of these measurements. Tuning the magnetic field $\mu_0 H\parallel c$ to 2.15 T, the previous single peak even splits into two
distinct peaks. While the lower-temperature peak in terms of shape and position seems to be related to the previous peak at lower magnetic fields (1.5 to 2.1 T), the upper one resembles the (single) peaks found at magnetic fields larger than 2.15 T. Increasing the magnetic field further, both transitions are suppressed to lower temperatures until above 2.5 T a gap-like behaviour of the specific heat arises.

  \begin{figure}[tb]
  \centering
  \includegraphics[width=\columnwidth]{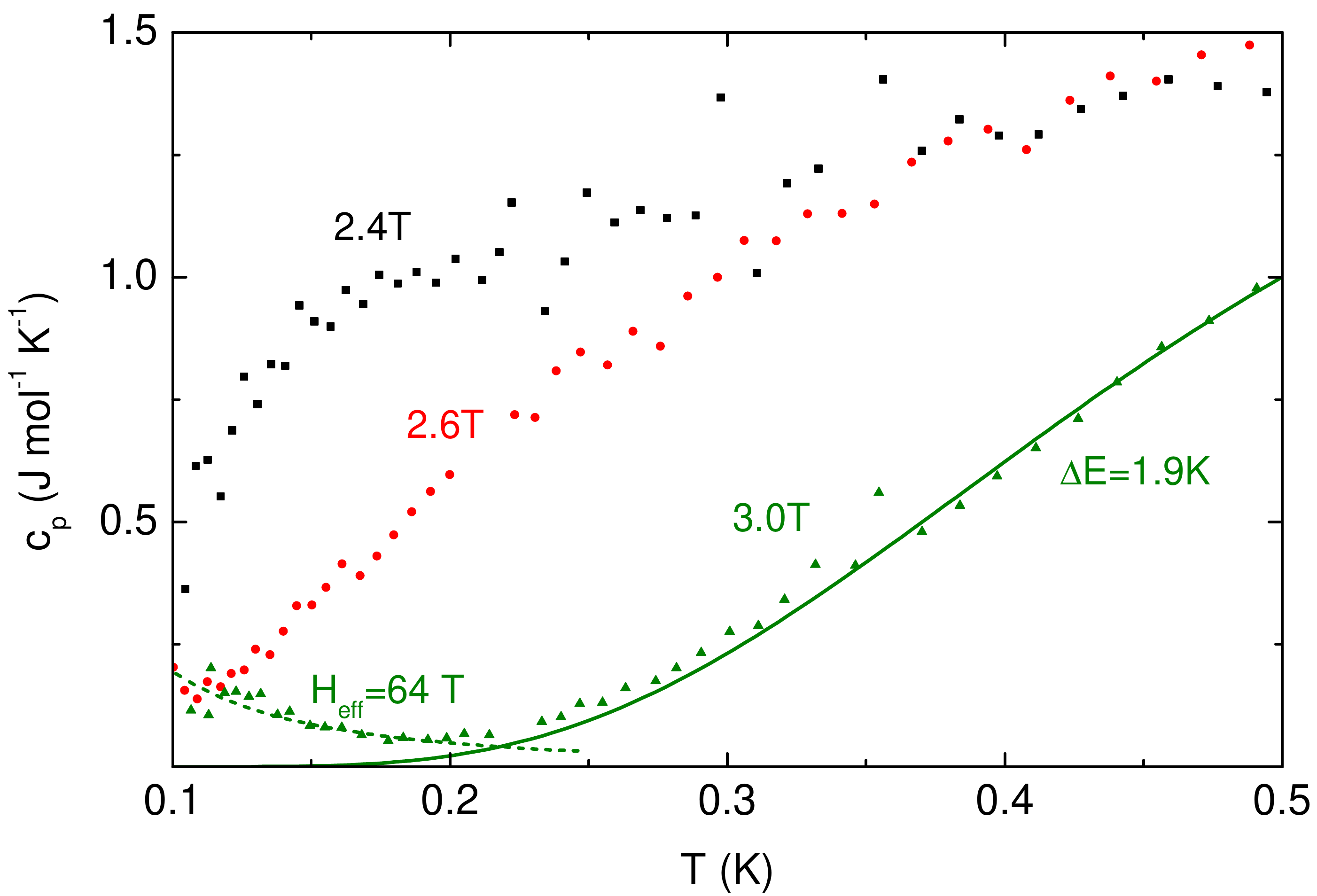}
  \caption{(Color online) Specific heat of Cs$_2$CoCl$_4$ in the gapped phase in magnetic fields applied along \textit{a}. The solid line represents a fit of a Schottky contribution given by Eq.~(\ref{eq:schottky}) and the dashed line is a fit of a nuclear contribution via Eq.~(\ref{eq:nuclear}).}
  \label{fig:cpHaHighFieldFit}
 \end{figure}
  The temperature dependence  in the high-field range $H\parallel\textit{a}$ is shown in fig.~\ref{fig:cpHaHighFieldFit} in more detail. With increasing field the specific heat is suppressed, corresponding to the opening of a gap by the magnetic field.
  At \unit[3.0]{T} a phenomenological description of the data is possible via a simple model of a two-level system with an energy gap $\Delta E$, yielding a Schottky contribution
  \begin{equation}
  \label{eq:schottky}
  \frac{c_p^{\Delta E}(T)}{R}= \left(\frac{\Delta E}{k_\textrm{B}T}\right)^2 \frac{e^{\Delta E/k_\text{B}T}}{\left(1+e^{\Delta E/k_\text{B}T}\right)^2}.
  \end{equation}
  A fit (solid line in fig.~\ref{fig:cpHaHighFieldFit}) of the specific heat $c_p^{\Delta E}(T)$ for the maximum field of 3~T yields the field-induced gap  $\Delta E/k_\text{B} \simeq \unit[1.9]{K}$. An additional low-temperature contribution is seen at $T\lesssim\unit[0.3]{K}$, which might stem from hyperfine interactions with the nuclear spin $I=7/2$ of Co, that lead to a specific heat contribution of the form
  \begin{equation}
  \label{eq:nuclear}
  \frac{c_p^N(T)}{R}=\frac{I+1}{3I} \frac{(\mu_N \mu_0 H_{\textrm{eff}})^2}{(k_\text{B} T)^2},
  \end{equation}
  with the induced effective field $H_\textrm{eff}$ and the nuclear magneton $\mu_N=e\hbar/2m_p$. Fitting $c_p^N$ to the low-temperature data yields an effective field $\mu_0 H_\textrm{eff}\simeq \unit[64]{T}$, which lies in the typical range obtained for metals, but exceeds the value reported for pure cobalt \cite{Lounasmaa1962,Dixon1965}.

 \begin{figure}[tb]
  \centering
  \includegraphics[width=\columnwidth]{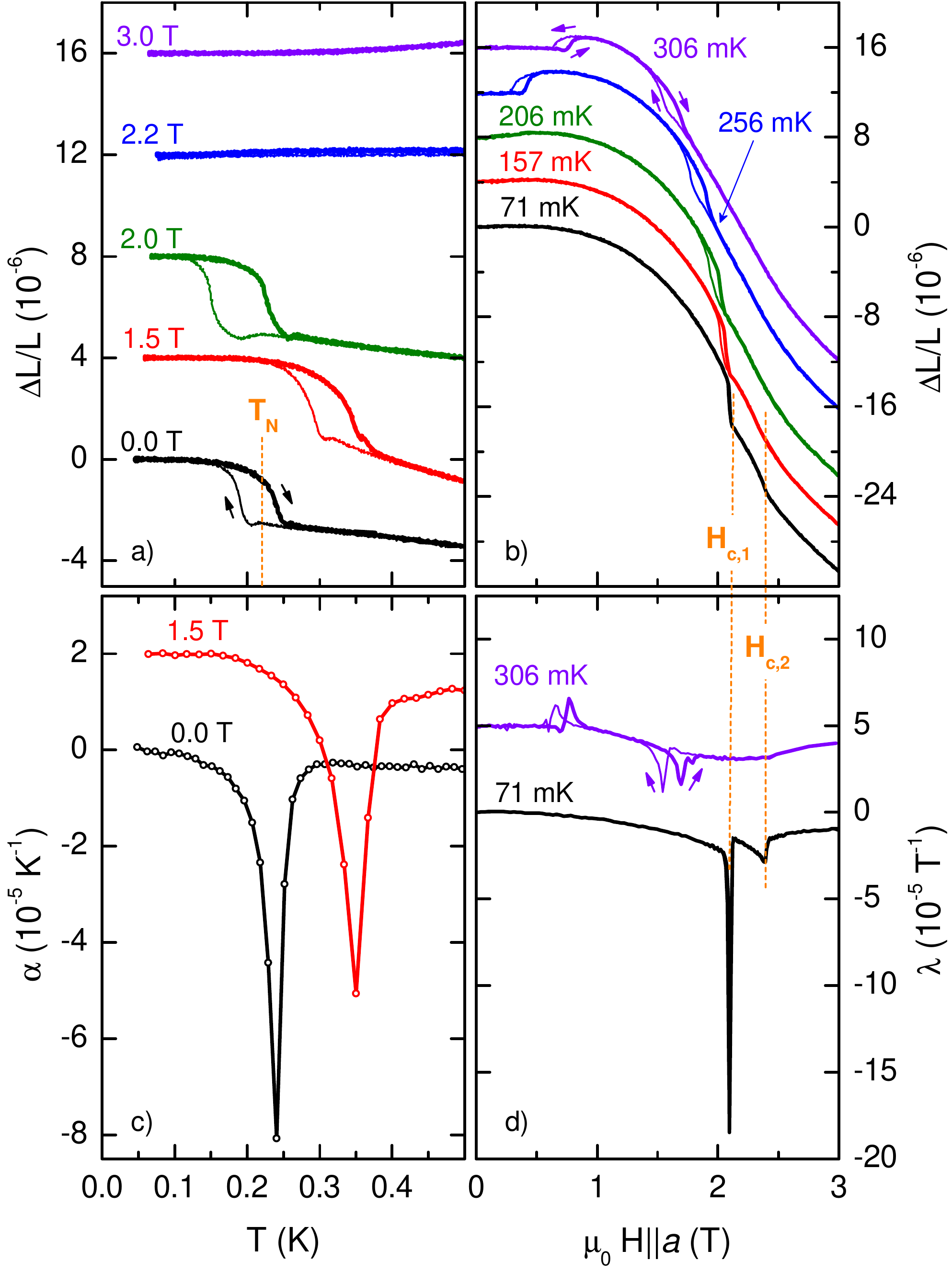}
  \caption{(Color online) Thermal expansion (left) and magnetostriction (right) of Cs$_2$CoCl$_4$ in magnetic fields applied along \textit{a}. For clarity, the $\Delta L/L$ curves in panels a) and b) are offset with respect to each other by $4\cdot 10^{-6}$ and the corresponding temperature and field derivatives $\alpha$ and $\lambda$ in panels c) and d) are offset by $2\cdot 10^{-6}$/K and $5\cdot 10^{-5}$/T, respectively.  Bold (thin) lines represent measurements with increasing (decreasing) temperature or field (indicated by arrows) with rates of 3~mK/min or 5~mT/min, respectively. In c), only $\alpha(T)$  obtained with increasing temperature are shown.}
  \label{fig:TADMSHa}
 \end{figure}
\begin{figure}[tb]
  \centering
  \includegraphics[width=\columnwidth]{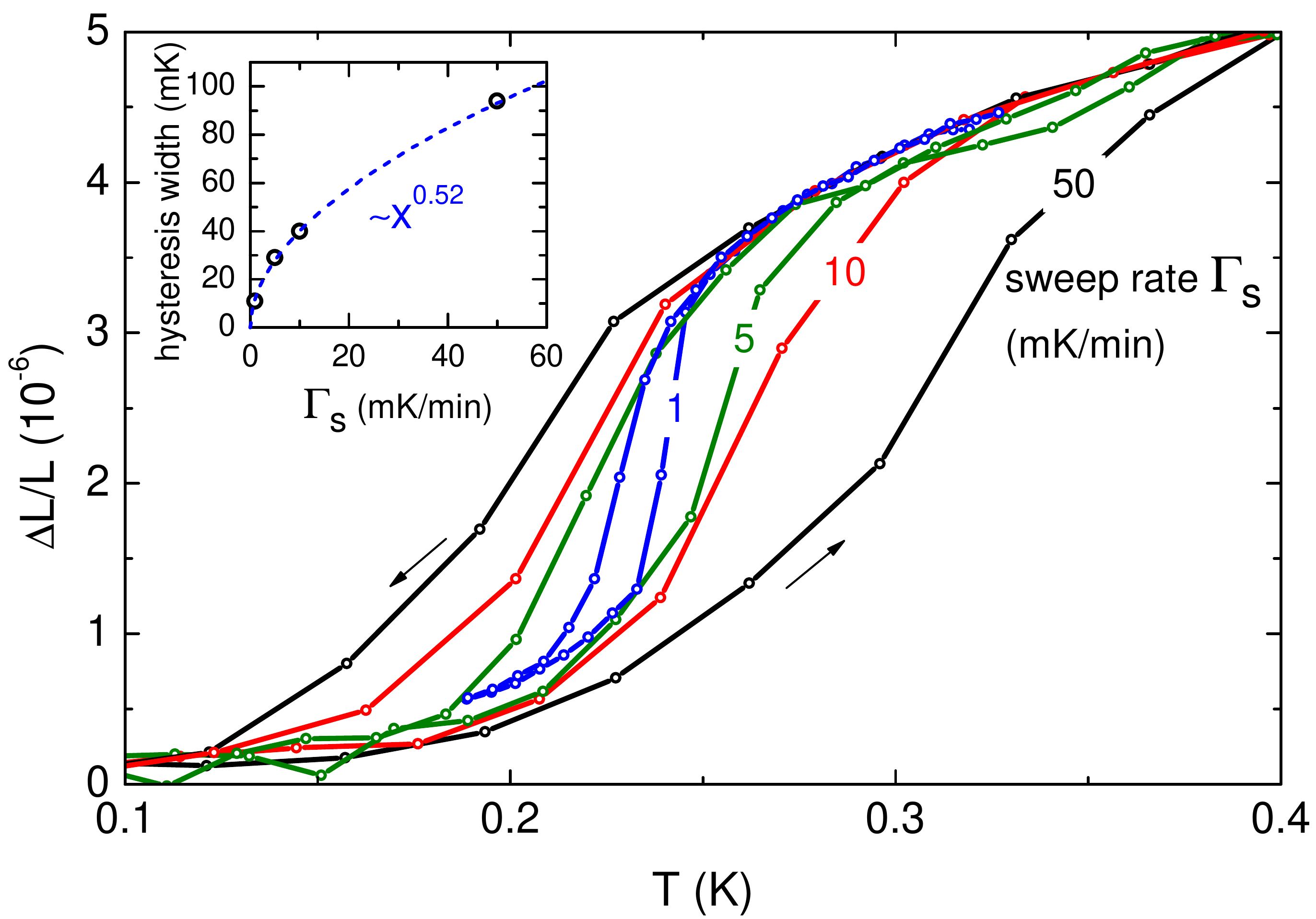}
  \caption{(Color online) Temperature hysteresis of the N\'{e}el transition  of Cs$_2$CoCl$_4$ in zero magnetic field as observed by measuring the thermal expansion $\Delta L(T)\parallel\textit{b}$ while  continuously increasing or decreasing the temperature (indicated by arrows) with different sweep rates. Inset: Dependence of the hysteresis width on the sweep rate. }
  \label{fig:RateDependence}
 \end{figure}

 Figure~\ref{fig:TADMSHa} displays the thermal expansion and magnetostriction $\Delta L/L\parallel a$ as a function of temperature and magnetic field. In general, we observe a field dependence analogous to that of the specific heat. With increasing temperature in zero magnetic field, $\Delta L/L$ drops by approximately $2\cdot10^{-6}$ at a temperature of $\unit[240]{mK}$, which is higher than the $T_\textrm{N}\simeq \unit[220]{mK}$ found in specific heat. However, the transition is accompanied by a strong temperature hysteresis, that is centered roughly around the $T_\text{N}$ extracted from specific heat (marked by a dashed line in fig.~\ref{fig:TADMSHa} a). Due to the experimental issues in the process of recording the single data points of the specific heat that were discussed above, hysteresis effects of $c_p$ are possibly reduced in comparison to the thermal expansion data, which are obtained continuously while slowly heating or cooling the sample. In order to further investigate the character of the antiferromagnetic transition, measurements were performed at different sweep rates from \unit[1]{mK/min} up to \unit[50]{mK/min}  with $\Delta L/L$ measured along \textit{b} (fig.~\ref{fig:RateDependence}). The relative length change sharpens by decreasing the sweep rate and resembles a step-like anomaly as typically expected for a first-order transition. However, the width of the associated temperature hysteresis does not remain finite in the limit of vanishing sweep rate. It can be described by a power-law dependence with an exponent of $0.52$, as fitted by the dashed line in the inset of fig.~\ref{fig:RateDependence}. Due to the strong thermal coupling of the sample we exclude experimental origins for the hysteresis narrowing, but ascribe it to the dynamics of domain walls at the phase transition. 
 The observed $\sqrt{\Gamma_s}$ dependence of the width of the transition, where $\Gamma_s$ is the sweep rate, can be explained by a simple model which assumes that (i) the transition is of first order and (ii) that the dynamics of the transition is dominated by the motion of domain walls (rather than by their nucleation). The difference of the free energy densities, $\Delta f$, of two phases is approximately linear in $T-T_c$ for a first order transition, $\Delta f\propto  T-T_c$. In addition, $\Delta f$ can directly be identified with the force per area on a domain wall as the free energy $\Delta F =\Delta f \, A \, \Delta r$ is gained when a domain wall with the area $A$ moves by the distance $\Delta r$. Assuming that the velocity of the domain wall is proportional to the force and using that $T-T_c=\pm \Gamma_s t$ for a sweep across the first-order transition, one finds that $\Delta r \propto \Gamma_s t^2$. The phase transition is completed at the time $t_s$, when $\Delta r$ is of the order of the distance of nucleation centers of domains and we find $t_s\propto 1/\sqrt{\Gamma_s}$. Therefore the hysteresis width can be expected to be of the order of $\Gamma_s t_s \propto \sqrt{\Gamma_s}$ as observed.
 Together with the latent heat observed in the specific heat raw data, we infer that the transition is of first order and mainly driven by the motion of domain walls.

 Maintaining this hysteretic character, the transition shifts to higher temperatures upon increasing the magnetic field to \unit[1.5]{T}, indicated by sharp peaks of $\alpha$ centered around $\unit[320]{mK}$. Similar to the heat-capacity results it then is suppressed. In contrast, no transition is observed at all in thermal expansion for magnetic fields larger than 2.1 T, whereas clear anomalies are present in heat capacity. The absence of features in $\alpha$ signals a vanishing pressure dependence of the transition temperature $T_C$, which linearly enters the thermal expansion\footnote{The uniaxial pressure dependence of the transition temperature $T_c$ relates to the uniaxial thermal expansion $\alpha_i$ in case of a 2\textsuperscript{nd}-order phase transition via the Ehrenfest relation $\frac{d T_c}{d p_i}=T_cV_m\frac{\Delta\alpha_i}{\Delta c}$ and to the length change $\Delta L_i$ in case of a 1\textsuperscript{st}-order transition via the Clausius-Clapeyron relation $\frac{dT_c}{dp_i}=\frac{\Delta L_i/L}{\Delta S/V_m}$}, $\alpha\propto\frac{\partial T_C}{\partial p}$. Thus, the absence of peaks in $\alpha$ does not exclude a phase transition, but only indicates that the corresponding transition has a negligible pressure dependence.

 At elevated temperatures ($T>\unit[250]{mK}$), the magnetostriction reveals two distinct anomalies  as a function of the magnetic field, which indicate transitions between the paramagnetic and the ordered phases. Due to a maximum of $T_\textrm{N}(B)$ close to \unit[1]{T}, this phase boundary can be passed twice as a function of the magnetic field. Both transitions are accompanied by a sizable hysteresis that in case of the upper transition monotonically decreases upon cooling and vanishes completely at the lowest measured temperature. Besides the very pronounced anomaly close to $H_{c,1}=\unit[2.1]{T}$ an additional kink is seen in $\Delta L/L$ at $H_{c,2}=\unit[2.4]{T}$ for the lowest temperature of \unit[71]{mK}, that is associated with a minimum in $\lambda$ (marked by dashed lines in \ref{fig:TADMSHa}). From 71 mK up to 157 mK this minimum remains at almost the same magnetic field. No hysteresis is observed at any of the measured temperatures for this transition, which indicates a second-order phase transition at $H_{c,2}$. 
\subsection{Magnetic field along b}

As described above a magnetic field along \textit{b} is unique for the actual crystal symmetry, because $b$ is the only principal axis that lies within both types of magnetic easy-plane orientations, which differ by a rotation $\pm\beta$ around \textit{b} (see fig.~\ref{fig:struktur}). With respect to the local coordinate system of the individual Co$^{2+}$ ions, this field configuration is therefore sufficiently described by a single component term $\propto H_bS_i^y$ in the Hamiltonian (\ref{eqn:HXXZ}). Moreover, as the spins in the zero-field N\'{e}el phase are oriented almost along \textit{b}, one may expect spin-flop transitions for this field direction.

\begin{figure}[tb]
  \centering
  \includegraphics[width=0.9\columnwidth]{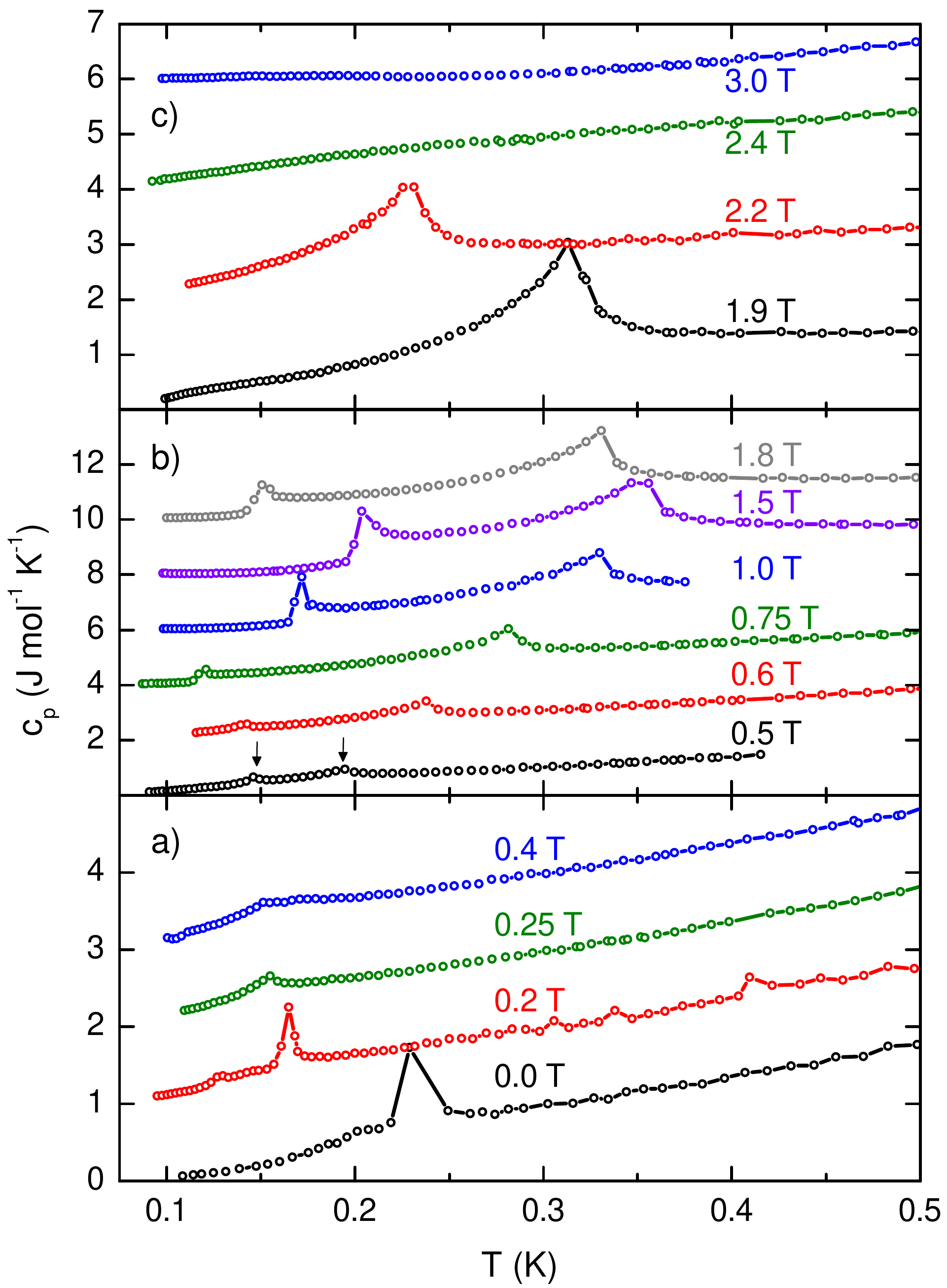}
  \caption{(Color online) Specific heat of Cs$_2$CoCl$_4$ for different magnetic fields $H\parallel\textit{b}$. For clarity, the curves are offset with respect to each other by 1~J mol$^{-1}$K$^{-1}$ in panel a) and by 2~J mol$^{-1}$K$^{-1}$ in panels b) and c). The arrows in b) indicate two distinct phase transitions, which are present in a field range from 0.5 to 1.8 T. }
  \label{fig:cpHb}
 \end{figure}

As is shown in fig.~\ref{fig:cpHb}, the obtained specific-heat data for $H\parallel\textit{b}$  significantly differ from those obtained in magnetic fields applied along \textit{a} or \textit{c}. While $T_{\text N}$ is continuously enhanced by small magnetic fields $H\parallel a$ and $H\parallel c$, magnetic fields $H\parallel b$ instead cause an initial suppression of the transition temperature. At \unit[0.2]{T} a small additional anomaly appears around $\unit[130]{mK}$, which, however, is not seen at \unit[0.25]{T}, where only one single peak is observed again. Applying magnetic fields above 0.4~T splits the single peak into two and the upper peak moves continuously to higher temperatures until it reaches a maximum value of \unit[0.35]{K} at an applied magnetic field of \unit[1.5]{T}. In contrast, the lower peak shows a non-monotonic field dependence. Starting at \unit[0.4]{T}, it first is suppressed to a minimum temperature of about $\unit[117]{mK}$ at \unit[0.75]{T}. Then, the peak position is shifted towards higher temperatures again and reaches a maximum of around $\unit[200]{mK}$ at a magnetic field of 1.5 T. Further increasing the magnetic field to \unit[1.8]{T} rapidly shifts the lower transition to lower temperatures until it is finally no longer observable within the experimental temperature range in a magnetic field of 1.9 T. The position of the upper peak is weakly field dependent in the field range between 1.5 T and 2.0 T, but then also is shifted towards $T\to0$ in magnetic fields above 2.2~T. In magnetic fields above 2.8~T an overall suppression and a gap-like behaviour of the specific heat indicates the opening of a gap by the magnetic field as expected for full spin polarization.

\begin{figure}[tb]
  \centering
  \includegraphics[width=\columnwidth]{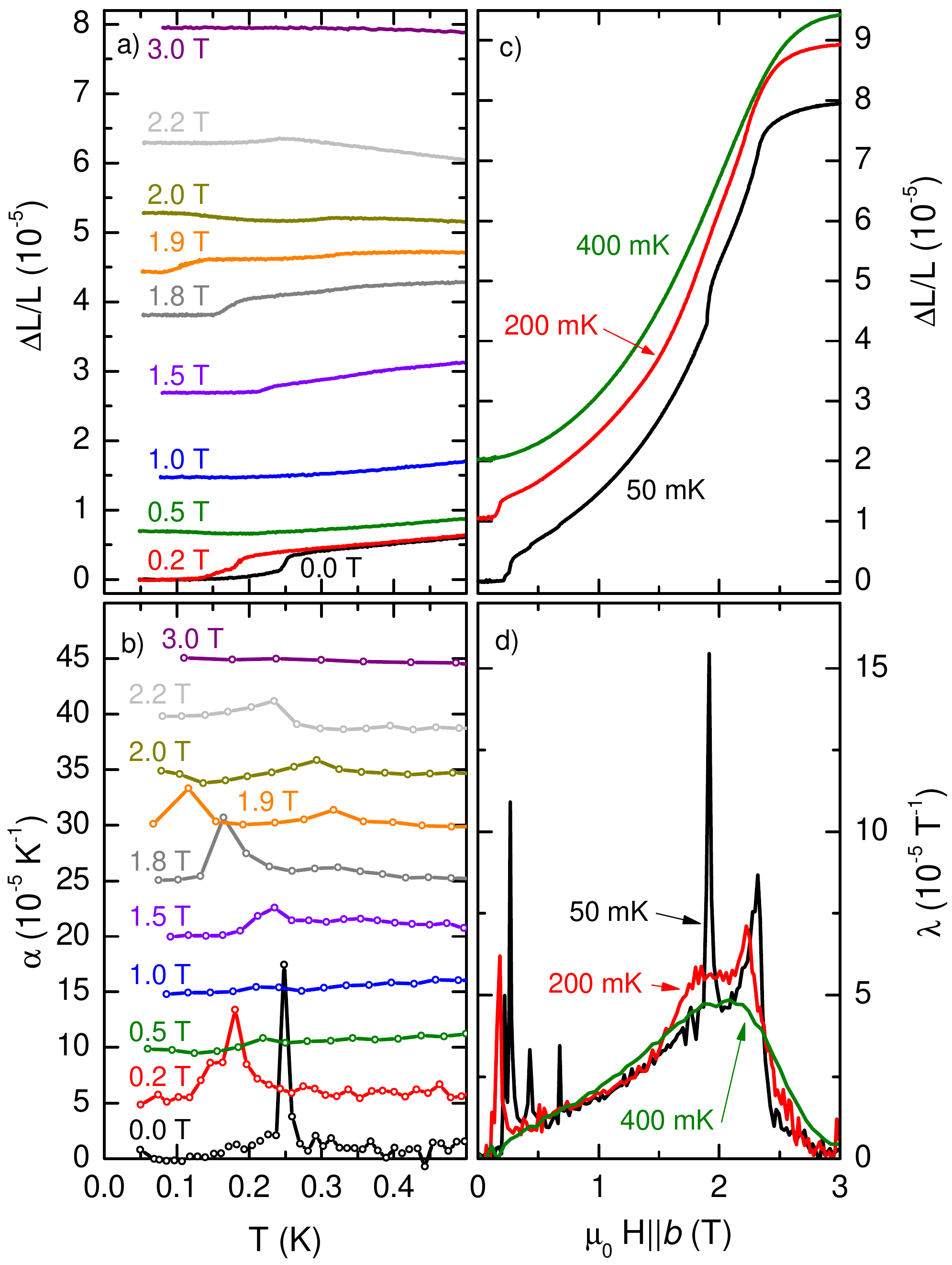}
    \caption{(Color online) Thermal expansion (left) and magnetostriction (right) of Cs$_2$CoCl$_4$ in magnetic fields along \textit{b}. The relative length changes $\Delta L/L$ are displayed in a) and c), while b) and d) show the corresponding temperature and field derivatives.  In a) the $\Delta L(T)/L$ curves for different magnetic fields are shifted according to the measured magnetostriction $\Delta L(H)/L$ at  $T=\unit[50]{mK}$ shown in c). For clarity, the other curves in c) are offset with respect to each other by $10^{-5}$ and the $\alpha(T)$ curves in b) by $5\cdot 10^{-5}$/K. In all panels, only data obtained with increasing temperature or field are shown.}
    \label{fig:TADMSHb}
 \end{figure}

A similarly rich magnetic-field dependence arises in the measurements of the thermal expansion and the magnetostriction (fig.~\ref{fig:TADMSHb}). In zero magnetic field, the relative length change $\Delta L(T)/L$ reveals a step-like anomaly close to $T_\text{N}$, that is shifted to lower temperatures by increasing the magnetic field and, as already discussed above, this transition is accompanied by a hysteresis depending on the temperature sweep rate. In the field range from \unit[0.5]{T} to $\unit[1.3]{T}$, the thermal expansion $\Delta L/L \parallel\textit{b}$ does not show comparably sharp anomalies as are seen in the specific heat. In magnetic fields above \unit[1.5]{T}, a kink of $\Delta L/L$ reappears close to \unit[0.2]{K} that is shifted towards lower temperatures by increasing the magnetic field. Moreover, at \unit[1.9]{T} another anomaly appears around 0.3~K, which then also shifts towards lower temperature when the field is further increased and finally vanishes around \unit[2.3]{T}. Clearer evidence of magnetic phase transitions are seen in the magnetostriction data. Here, a saturation of $\Delta L/L$ sets in at magnetic fields $\mu_0 H>\unit[2.3]{T}$, similar to the magnetization, which for $H\parallel\textit{a}$ is known to be saturated in this field range as well \cite{Kenzelmann2002}. At the lowest temperature of \unit[50]{mK}, two pronounced anomalies are present at $\unit[0.26]{T}$ and $\unit[1.9]{T}$, which are reflected by sharp peaks of $\lambda$, see fig.~\ref{fig:TADMSHb}~d). Moreover, additional smaller anomalies occur at $0.22$, $0.43$, and $\unit[0.67]{T}$, which can be no longer resolved when the temperature is increased to \unit[200]{mK}. Concerning the larger anomalies, the upper two are strongly broadened at \unit[200]{mK}, but still can be seen as maxima in $\lambda$, while the lower one is only weakly broadened and shifted to $\unit[0.18]{T}$. At \unit[400]{mK}, $\lambda(H)$ displays only one broad maximum around $\unit[2]{T}$, which signals the one-dimensional magnetism in the paramagnetic phase. No signatures of sizable field-induced hysteresis are found at any temperature in the magnetostriction data.

\subsection{Phase diagrams}
  \begin{figure}[t]
  \centering
  \includegraphics[width=\columnwidth]{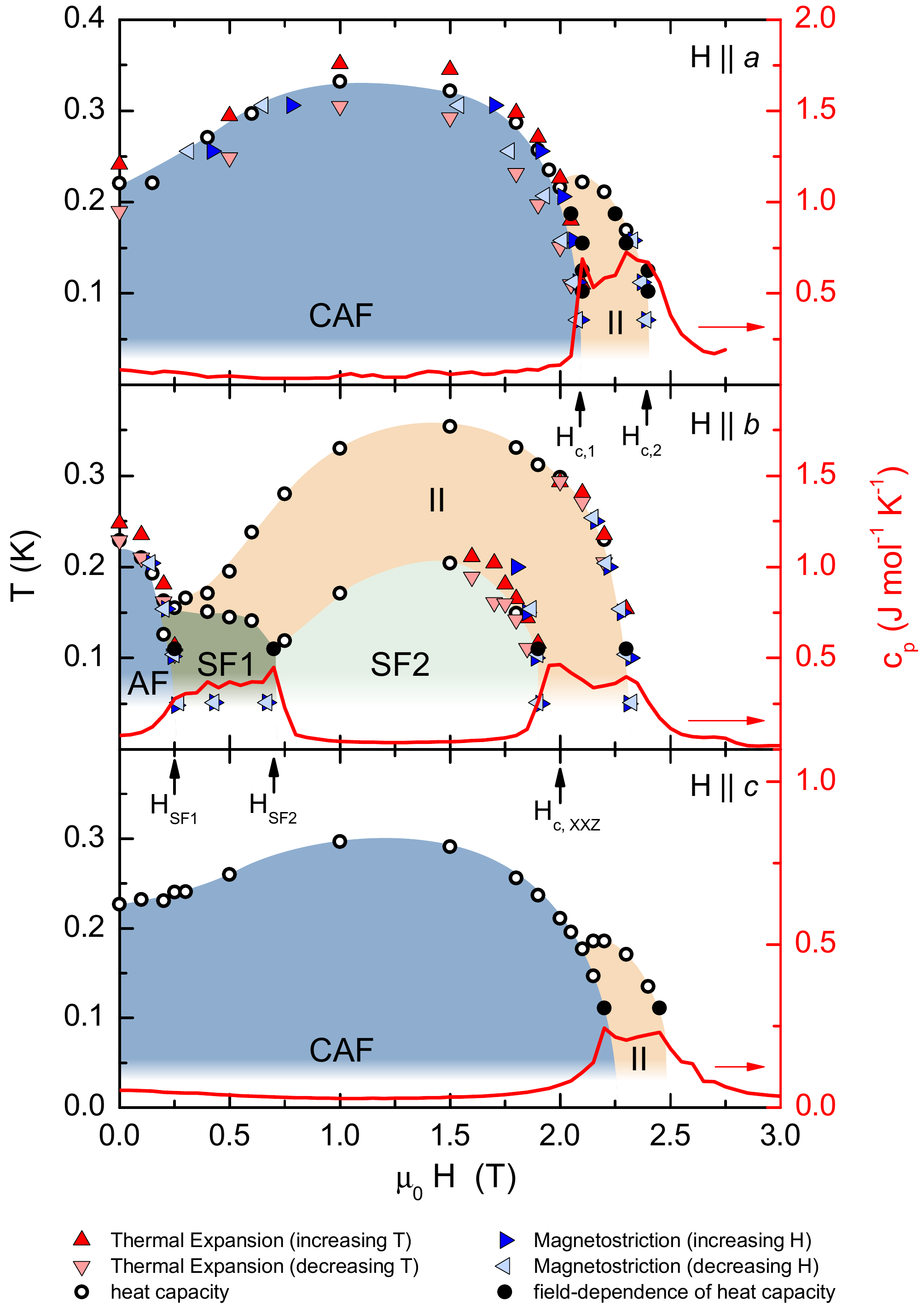}
  \caption{(Color online) Low-temperature $H-T$ phase diagrams of Cs$_2$CoCl$_4$ for magnetic fields along all three crystallographic axes. Shaded areas are guides to the eye. The magnetic-field dependent specific heat $c_p$ at a constant temperature of \unit[0.11]{K} is also displayed (red lines, right scales). For $H\| b$, the occurrence of two spin-flop phases, SF1 and SF2, is indicated by $c_p(H)$ and for all three field directions $c_p(H)$ is strongly enhanced in the field range of phase~II.}
  \label{fig:Phasediagrams}
 \end{figure}
All phase transitions observed by the specific-heat and thermal-expansion measurements are plotted in fig.~\ref{fig:Phasediagrams} in field versus temperature phase diagrams for the three principal magnetic-field directions discussed before. The data points of different methods agree within the bounds of hysteresis effects. In general, the phase diagrams are very similar for magnetic fields $H\parallel$ \textit{a} and \textit{c}. A slight rescaling of the field axis from \textit{a} to \textit{c} is probably induced by a small $g$-tensor anisotropy. We find an initial increase of the zero-field N\'{e}el temperature $T_\text{N}\simeq \unit[220]{mK}$ by small magnetic fields up to \unit[1.5]{T} and a subsequent suppression of $T_\text{N}\rightarrow 0$, which is accompanied by a decreasing hysteresis. Combining the phase transitions observed by different experimental methods, the presence of a previously unknown and well defined low-temperature phase adjacent to the antiferromagnetic phase becomes obvious, which is
referred to as ``phase II'' in the following. The entry of phase II is accompanied by a steep increase of the heat capacity as a function of the magnetic field in-between the critical fields $\mu_0 H_{c,1}\simeq \unit[2.1]{T}$ and $\mu_0 H_{c,2}\simeq\unit[2.4]{T}$ (right scale of fig.~\ref{fig:Phasediagrams}). In previous studies by neutron diffraction~\cite{Kenzelmann2002} a coinciding critical field $H_{c,1}$ is found bordering the antiferromagnetically ordered state from a state that due to the lack of antiferromagnetic order reflections has been argued to be a spin-liquid state. In case of a spin-liquid ground state one would expect a continuous temperature evolution of thermodynamic properties with no signs of magnetic ordering down to zero temperature. Our data, however, give clear evidence of temperature-dependent ordering transitions in the magnetic field range $H_{c,1}<H<H_{c,2}$. Thus, our thermodynamic data suggest the occurrence of another type of order in this field range instead of a disordered spin-liquid ground state in Cs$_2$CoCl$_4$.

Applying magnetic fields along \textit{b} leads to a more complex phase diagram. Here, the formerly discussed phase~II similarily arises as a function of magnetic field at lowest temperatures. At elevated temperatures phase~II, however, extends over a wide field range and even merges with the antiferromagnetic and the spin-flop phases (discussed later) at small magnetic fields in a triple point. The extension of phase~II explains the appearance of two distinct transitions as a function of temperature in the field range from 0.5 to \unit[1.8]{T} as observed in the specific heat data. The enhanced specific heat in the field range of phase~II as well as the power-law scaling $c_p\propto T^\alpha$ are almost identical for all three field directions. The exponents $\alpha$ obtained by fitting the temperature dependences of $c_p$ within phase~II agree within a few percent (see table~\ref{tab:powerlaw}). Thus, we conclude that phase~II is present irrespective of the magnetic field direction, but is most favored by a magnetic field applied along \textit{b}.

\begin{figure}[t]
  \centering
  \includegraphics[width=\columnwidth]{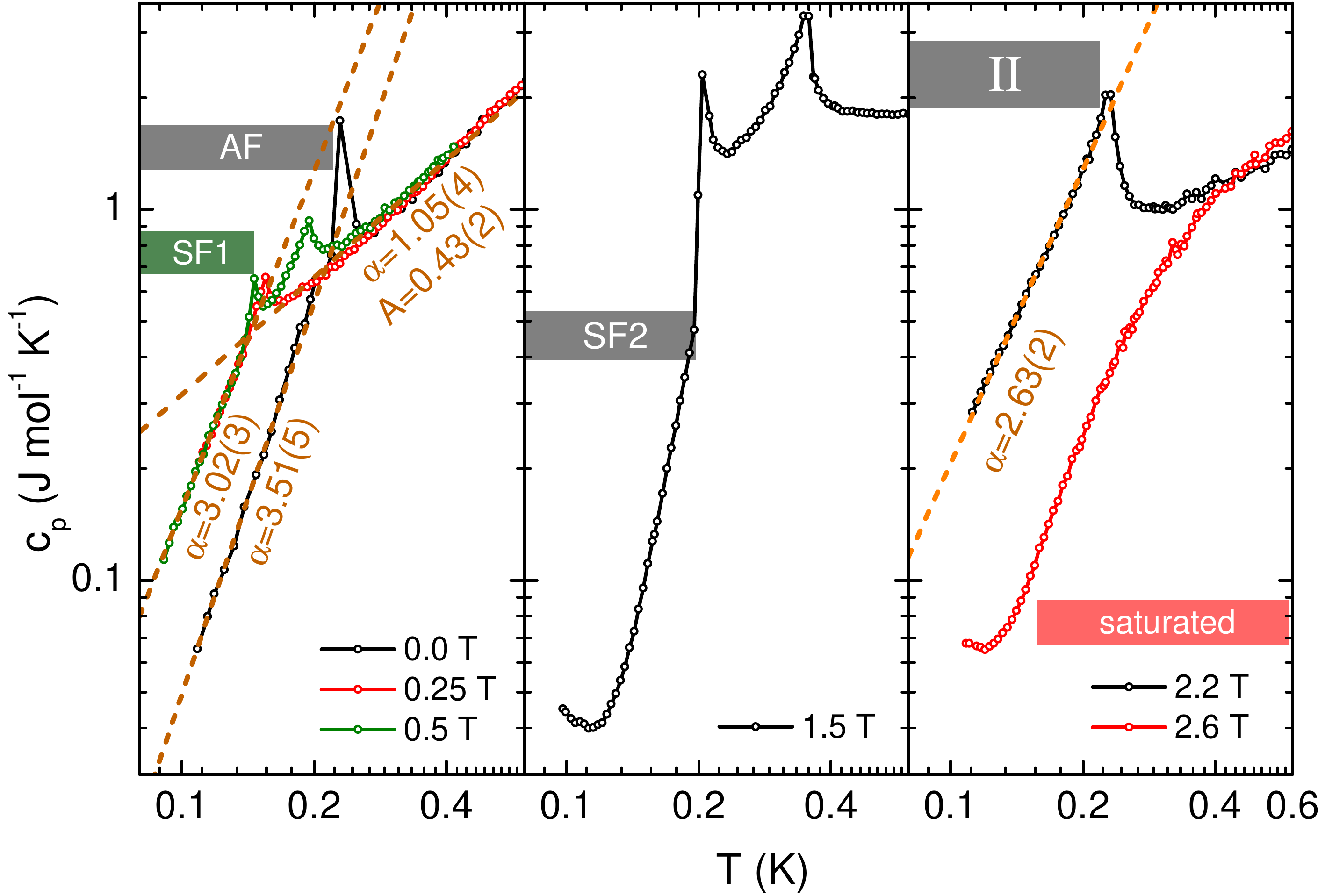}
  \caption{(Color online) Representative temperature dependences of the specific heat of the various  low-temperature phases of Cs$_2$CoCl$_4$. Shaded labels indicate the respective low-temperature phase at the given magnetic field applied along \textit{b}. Dashed lines are fits of $c_p(T)/R=A\cdot T^\alpha$.}
  \label{fig:cpHbScaling}
 \end{figure}
 \begin{table}[tb]
   \begin{ruledtabular}
\begin{tabular}{llrrr}
\toprule
     &               &  $\alpha$ && \\
 Phase &  & $H||a $& $H||b$ & $H||c$\\
 \hline
AF &($<\unit[0.25]{T}$)& 3.5 & 3.5 & 3.5\\
SF1 &(0.25-\unit[0.7]{T})& - & 3.02 & -\\
SF2 & (0.7-\unit[2]{T})& - & complex & -\\
II &($\approx$2-\unit[2.4]{T})& 2.68(15) & 2.63(4) & 2.69(7)\\
\bottomrule
 \end{tabular}
 \end{ruledtabular}
\caption{Power-law exponents $\alpha$ of the specific heat $c_p\propto T^\alpha$ within the various  magnetic phases for different magnetic field directions. The phases SF1 and SF2 only occur for $H\parallel\textit{b}$.}
\label{tab:powerlaw}
\end{table}

While phase II arises for all three magnetic field direction, the low-field phases differ in case of $H\parallel\textit{b}$. As described above~\cite{Kenzelmann2002}, the spins in the zero-field N\'{e}el phase are tilted away from \textit{b} by $\phi \approx \pm 15^\circ$. Neglecting these tilts makes the configurations $H\parallel \textit{a}$ and $H\parallel\textit{c}$ symmetry equivalent. As in both cases the magnetic field points almost perpendicular to the ordered moments, it causes a canting of the spins towards the field direction and the corresponding phases are denoted as canted antiferromagnet (CAF). Magnetic fields $H\parallel\textit{b}$, in contrast, have a large component collinear to the ordered moments and typically induce spin-flop transitions. In fact, our data indicate a field-induced transition at a small magnetic field $\mu_0 H_{SF1}\simeq\unit[0.25]{T}\parallel b$ that is indicated by a rise of $c_p$ by a factor of about $5$ and a peak in the magnetostriction coefficient $\lambda$. For $H<H_{SF1}$, we find the same temperature dependence $c_p\propto T^{3.5}$ as for the other magnetic field directions up to about \unit[0.5]{T}, but above $H_{SF1}$ the exponent $\alpha$ is reduced to 3.0 and remains constant up to $\mu_0 H_{SF2}\simeq\unit[0.7]{T}$, above which $c_p$ is strongly reduced again and is no longer described by a simple power; see figs.~\ref{fig:Phasediagrams}, \ref{fig:cpHbScaling} and table~\ref{tab:powerlaw}. Both transition fields, $H_{SF1}$ and $H_{SF2}$, are also seen in the magnetostriction data at the lowest temperature of $50~\text{mK}$ as peaks in $\lambda(H)$, see fig.~\ref{fig:TADMSHb}~d). Interestingly, the magnetostriction data even indicate additional transitions slightly below $H_{SF1}$ and $H_{SF2}$, but because these anomalies weaken above 100~mK, it is not clear, at present, whether they might be related to some domain effects. 

\begin{figure}[b]
  \centering
  \includegraphics[width=\columnwidth]{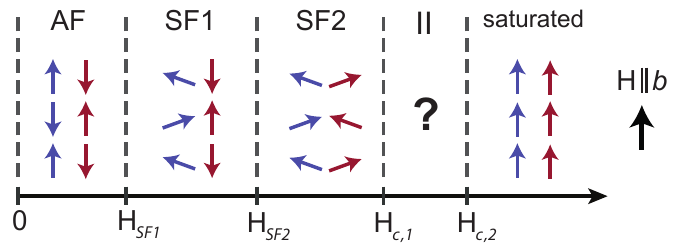}
  \caption{(Color online) 
  Sketch of the proposed sequence of field-induced phases in Cs$_2$CoCl$_4$ for 
H$\parallel b$. The different colors of neighboring spin chains represent their different easy-plane orientations. AF: all spins are nearly \mbox{(anti-)parallel} to the field H and to each other. SF1: every second chain transforms to a spin-flop phase, which is stabilized by Dzyaloshinskii-Moriya interactions between spins of neighboring chains. SF2: spin-flop phase in all chains. Phase II could be an incommensurate or nematic phase preceeding the fully saturated phase. 
  }
  \label{fig:spinflop}
 \end{figure}

We speculate that the two transitions at $H_{SF1}$ and $H_{SF2}$ signal a two-stage spin-flop transition, which may arise from DM interactions between spins of neighboring chains. In Cs$_2$CoCl$_4$, the symmetry of the corresponding Co-Cl-Cl-Co bonds is indeed low enough to allow for interchain DM interactions, which favor a perpendicular alignment of the spins of neighboring chains. Thus for $H\| b$, a spin configuration with a spin-flop state in every second chain, see fig.~\ref{fig:spinflop}, may be stabilized by the energy gain due to interchain DM interactions in the intermediate field range $H_{SF1}<H<H_{SF2}$, until above $H_{SF2}$ the fully spin-flopped state evolves. Such two-stage spin-flop transitions were previously observed, \textit{e.g.}, in BaCu$_2$Si$_2$O$_7$ \cite{Tsukada2001}.

\section{Summary and Conclusions}

Our measurements of the specific heat and thermal expansion of Cs$_2$CoCl$_4$ reveal a rich low-temperature phase diagram. Depending on the direction of the magnetic field, up to four differently ordered phases arise until above about 2.4~T the fully saturated state with a finite spin gap is reached. For $H\| a$ or $c$, \textit{i.e.}, for magnetic fields applied (almost) perpendicular to the spin orientations in the zero-field N\'{e}el state, the expected canted antiferromagnetic state is  present only up to about 2~T, while in the adjacent field range $H_{c1}< H < H_{c2}$ another low-temperature phase II occurs. Such a phase II also occurs for $H\| b$, for magnetic fields applied (almost) parallel to the spin orientations in the zero-field N\'{e}el state, and for this longitudinal field direction the phase II even extends over a much larger field and temperature range. In addition, our data identify two other low-temperature phases SF1 and SF2, which probably result from a two-stage spin-flop transition due to finite interchain DM interactions.

A main open question concerns the nature of the phase II. In the following, we would like to discuss three options: (i) a spin-liquid phase~\cite{Kenzelmann2002}, (ii) an incommensurate phase similar to the one observed in Cs$_2$CuCl$_4$ \cite{Veillette2005,Starykh2010} and (iii) a more exotic nematic phase.

Based on neutron scattering data for $H\| a$, which did not find any signature of magnetic order in the  $(b,c)$ plane, Kenzelmann \textit{et al.}
suggested that phase II is a spin-liquid state~\cite{Kenzelmann2002}, which one would expect to continuously evolve from the high-temperature disordered state. Our data, in contrast, reveal  clear anomalies as a function of temperature associated with a thermodynamic phase transition. While for some spin-liquid states finite-temperature phase transitions of emergent gauge degrees of freedom are possible, see \textit{e.g.} Ref.~\onlinecite{Nasu2014}, a more likely scenario is the existence of an unidentified ordered phase with broken symmetry. 

As shown by an extensive analysis for the compound Cs$_2$CuCl$_4$ in Ref.~\onlinecite{Starykh2010}, the natural candidate for a high-field phase is an incommensurate magnetic state. Cs$_2$CuCl$_4$ and Cs$_2$CoCl$_4$ have  the same orthorhombic crystal structure with space group $Pnma$ with the same type of magnetic frustration. The important difference of both magnetic systems arises, however, from the fact that the Cu$^{2+}$ ions represent almost isotropic Heisenberg spins, while the actual Co$^{2+}$ ions result in an effective XY-like spin-1/2 system with magnetic easy planes whose orientations alternate from chain to chain, see fig.~\ref{fig:struktur}. Nevertheless, due to the identical symmetry one can directly follow the symmetry-based arguments of Ref.~\onlinecite{Starykh2010} concerning the magnon spectrum for high magnetic fields.
An unavoidable consequence of the frustrated interchain coupling $J_{bc}$ is to shift the minima of the gapped magnon excitations at large magnetic field from momentum $\pi$ to incommensurate values. While frustrated coupling does not benefit from antiferromagnetic ordering, it {\em does} benefit from ordering at incommensurate momentum, {\textit{i.e.}, if neighboring spins along the chains tilt by an angle different from $\pi$. Thus, phase II, which for any field direction emerges in our phase diagrams upon reducing the magnetic field from the fully saturated phase at sufficiently low temperature, may be thought of as a condensate of magnons at incommensurate momentum. Furthermore, DM interactions also favor incommensurate order \cite{Starykh2010} at high fields. At  lower fields, in contrast, incommensurate order is suppressed by the effective Ising anisotropy arising from the coupling of spins with tilted easy planes, as has been already pointed out in Ref.~\onlinecite{Kenzelmann2002}. The theoretical analysis of Ref.~\onlinecite{Starykh2010} (including the effects of DM interactions) and the experiments on Cs$_2$CuCl$_4$ \cite{Veillette2005} both suggest that the incommensurate ordering vector in Cs$_2$CoCl$_4$ is oriented along the $b$ direction as well, in apparent conflict with the neutron scattering results~\cite{Kenzelmann2002} which did not find any ordering signal in phase II within the $(b,c)$ plane. Therefore either the incommensurate ordering vector is tilted out of the  $(b,c)$ plane by a mechanism not yet identified or another phase intervenes.

The absence of incommensurate order in this field range could give rise to a more exotic explanation of phase II, in terms of a nematic phase by a mechanism which was previously suggested to explain the 
nematic state in the pnictides, see  Ref.~\onlinecite{Fernandes2014}. The argument builds on the frustration inherent to the crystal structure.
By symmetry, the (commensurate) magnetization in the chains of type (1 \& 2), denoted by $M_A$,
and the magnetization in the chains of type (3 \& 4), denoted by $M_B$, is frustrated and thus in the free energy a coupling term $M_A M_B$ is forbidden by symmetry. Only a term of the form $(M_A M_B)^2$ is allowed by symmetry. Nematic ordering lifts this frustration:  while $\langle M_A \rangle =\langle M_B \rangle =0$, the nematic state is characterized by $\langle M_A M_B \rangle \ne 0$ and therefore induces with the help of the $(M_A M_B)^2$ term a linear coupling of the two subsystems. As has been shown for the pnictides\cite{Fernandes2014}, such a nematic transition is triggered when the correlation length in the unfrustrated subsystems is sufficiently long. Within this scenario a nematic phase would intervene in all cases where the transition to the low-temperature, low-field phase is of second order. This is consistent with our observations: all direct transitions from the paramagnetic to the AF (or CAF) phase appear to be of first order.

Previous studies\cite{Breunig2013} of the thermodynamics of Cs$_2$CoCl$_4$ in the temperature and field range where the magnetism is governed by only one dominant exchange coupling revealed signatures of quantum criticality that scale linearly as a function of magnetic field and extrapolate to a quantum critical field $H_{c,\text{XXZ}}\simeq\unit[2.0]{T}$ . This value is indicated by an arrow in fig.~\ref{fig:Phasediagrams}. It does not coincide with one of the extrapolated low-temperature phase boundaries, but rather is located within phase II. Irrespective of the microscopic origin, the formation of an extra phase as a consequence of competing energy scales is a well established phenomenon known for example from different families of unconventional superconductors. In analogy to the strong magnetic fluctuations present in those compounds, in Cs$_2$CoCl$_4$ strong quantum fluctuations are induced by the applied magnetic field which has non-commuting components for all spatial directions of the field due to the particular orientations of the magnetic easy planes.

With regard to future work, a deeper understanding of the precise nature of phase II in our phase diagram is probably most desirable. A simple scenario involving ordering at incommensurate momentum was not found by Kenzelmann \textit{et al.} \cite{Kenzelmann2002} in  neutron scattering. This opens the possibility of a more exotic nematic phase with correlated fluctuations between its frustratedly coupled subsystems, which asks for an extended analysis by microscopic methods.

This work was supported by the Deutsche Forschungsgemeinschaft via SFB 608 and FOR 960.

\bibliographystyle{apsrev4-1}

\end{document}